\documentclass[twocolumn,prd,showpacs,nofootinbib,amsmath,amssymb]{revtex4}
\usepackage{graphicx}

\providecommand{\Real}{\mathop\mathrm{Re}\nolimits}

\renewcommand{\l}{\left(}
\renewcommand{\r}{\right)}

\begin{document}

\title{Pinning down the kaon form factors in
$K^+\to\mu^+\nu_\mu\gamma$ decay}
\author{F.~L.~Bezrukov}
  \email{fedor@ms2.inr.ac.ru}
\author{D.~S.~Gorbunov}
  \email{gorby@ms2.inr.ac.ru}
\author{Yu.~G.~Kudenko}
  \email{kudenko@wocup.inr.troitsk.ru}
\affiliation{
    Institute for Nuclear Research of the Russian Academy of Sciences,\\
    60th October Anniversary prospect 7a, Moscow 117312, Russia
}

\begin{abstract}
  We find that the normal muon polarization in the decay
  $K\to\mu\nu_\mu\gamma$  is very sensitive to the values of the kaon
  vector $F_V$ and axial-vector $F_A$ form factors.
  It is shown that the ongoing KEK-E246 experiment can
  definitely determine the signs of the sum of the   form factors   
    if their
  difference is fixed from other considerations. 
This method can also verify the form factor
  values and signs obtained from the  $K^+\to l^+\nu_le^+e^-$ decays.
   A  new  experiment with  
sensitivity to the  normal and
  transverse muon polarizations  of about $10^{-4}$ will provide a
  unique possibility to determine the   $F_V$ and
  $F_A$ values with a few percent accuracy.
\end{abstract}     


\maketitle


Quantum  
Chromodynamics (QCD) describes strong processes at high
energies with remarkable precision. Processes at low energies are usually
described  by the effective Lagrangian  
of the light mesons. The
matching of the corresponding coupling constants is very involved and
is often replaced   
simply by the requirement of reasonable 
fitting of  
low-energy experimental data. 

In this paper, we study the normal muon polarization in the decay
$K^+\to\mu^+\nu_\mu\gamma$~($K_{\mu2\gamma}$) 
in order to investigate the
possibility of extracting the values of the vector and axial-vector kaon form factors,
$F_V$ and $F_A$. 
Because of the lack of understanding of the QCD low-energy structure, there is no
definite prediction for the values of the $F_V$ and $F_A$ form
factors: the calculation of them is a model-dependent 
 procedure. 
So, the measurement of these form factors would  
provide a possibility to  
select among various candidates for the correct description of the QCD 
low-energy limit.

Introducing three unit vectors
\begin{equation*}
  \vec{e}_L= \frac{\vec{p}_\mu }{ |\vec{p}_\mu|} \,,\;
  \vec{e}_N= \frac{\vec{p}_\mu\times (\vec{q}\times \vec{p}_\mu)
    }{ |\vec{p}_\mu\times (\vec{q}\times \vec{p}_\mu)|} \,,\;
  \vec{e}_T = \frac{\vec{q}\times \vec{p}_\mu}{
    |\vec{q}\times \vec{p}_\mu|} \,,
\end{equation*}
with $p_\mu$ and $q$ being the four-momenta of $\mu^+$ and $\gamma$,
respectively, one can define the longitudinal ($P_L$), normal ($P_N$) and
transverse ($P_T$) components of the muon polarization as the
corresponding contributions to the squared matrix element of
the $K_{\mu2\gamma}$ decay,
\begin{equation*}
  |M|^2=\rho_{0}[1\!+\!(P_L {\vec e}_L+P_N {\vec e}_N+P_T {\vec
  e}_T)\!\cdot\! \vec{\xi}\,]\;,
\end{equation*}
with ${\vec \xi}$ being a unit vector along the  
muon spin and $\rho_{0}$
is 
\begin{multline*}
  \rho_0(x,y) = \frac{1}{2}e^2 G^2_F |V_{us}|^2 
    (1-\lambda) \times\\
  \left\{ 
  f_\mathrm{IB}(x,y)+f_\mathrm{SD}(x,y)+f_\mathrm{IBSD}(x,y) \right\} \;,
\end{multline*}
where the internal bremsstrahlung ($\mathrm{IB}$), structure 
dependent ($\mathrm{SD}$)
and interference contributions ($\mathrm{IBSD}$) are given as 
follows~\cite{Gabrielli:1992dp,Bijnens:1992en,Chen:1997gf} 
\begin{align}
  \label{fIB}
  f_\mathrm{IB}&=\frac{4m^2_\mu|f_K|^2}{ \lambda x^2}
    \left[x^2+2(1-r_\mu)\left(1-x-\frac{r_\mu}{\lambda}\right)\right]\!,
  \\
  \label{fSD}
  f_\mathrm{SD}&= m^4_K x^2\bigg[|F_V+F_A|^2\frac{\lambda^2 }{ 1-\lambda}
      \left(1-x-\frac{r_\mu}{ \lambda}\right) \nonumber \\
    &\qquad+|F_V-F_A|^2(y-\lambda)\bigg]\;,
  \\
  \label{fIBSD}
  f_\mathrm{IBSD}&= - 4m_K m^2_\mu\bigg[\Real[f_K(F_V+F_A)^*]
      \left(1-x-\frac{r_\mu}{\lambda}\right) \nonumber\\
    &\qquad -\Real[f_K(F_V-F_A)^*]\frac{1-y+\lambda}{\lambda}\bigg] \;.
\end{align}
Here we used the standard notations $\lambda=(x+y-1-r_\mu)/x$,
$r_\mu=m^2_\mu/m^2_K$, and $x=2E_{\gamma}/m_K$, $y=2E_\mu/m_K$ with
$E_\gamma$, $E_\mu$ being the photon and muon energies in the kaon
rest frame, respectively; $G_F$ is the Fermi constant, $V_{us}$ is the
corresponding  
element of the Cabibbo--Kobayashi--Maskawa (CKM) matrix, 
$f_K$ is the kaon decay constant.   In terms of these variables the
differential decay width reads
\begin{equation*}
  d\Gamma(\vec{\xi}\,) =
    \frac{m_K}{32(2\pi)^3}|M(x,y,\vec{\xi}\,)|^2 dx\,dy \;.
\end{equation*}
The normal muon polarization $P_N$ is equal to the following asymmetry in
the partial decay width
\begin{equation}\label{PT}
  P_N=\frac{d\Gamma(\vec{e}_N)-d\Gamma(-\vec{e}_N) }
           {d\Gamma(\vec{e}_N)+d\Gamma(-\vec{e}_N)}
     \equiv\frac{\rho_N}{\rho_0} \;, 
\end{equation}
and at the tree level one has~\cite{Gabrielli:1992dp,Bijnens:1992en,Chen:1997gf}
\newpage 
\begin{widetext} 
\begin{multline}
  \rho_N(x,y) =
  e^2G_F^2|V_{us}|^2\frac{(1-\lambda)
  \sqrt{\lambda y-\lambda^2-r_\mu}}{m_K\lambda \sqrt{y^2-4r_\mu}}
  \Biggl\{\frac{4m_\mu^3|f_K|^2}{\lambda x}(x+y-2\lambda)
  -m_K^4m_\mu\lambda x^2\biggl[ |F_V+F_A|^2\frac{\lambda}{
  1-\lambda}\l 1-x-\frac{r_\mu}{\lambda}\r\\ +|F_V-F_A|^2
  \biggr]
  -2m_K^3m_\mu\left[\Real[f_{K}(F_V+F_A)^{*}]\l
  \frac{(r_\mu-\lambda )(1-x-r_\mu)}{1-\lambda}+\lambda x(1-x)\r
  -\Real[f_{K}(F_V-F_A)^*](y-2r_\mu)\right]\Biggr\}\;.
\nonumber  
\end{multline}
\end{widetext}
Since in the Standard Model both $\rho_0$ and $\rho_N$ are of the same
order, $P_N$ is of order one.

Only the absolute values of the sum and difference
of the kaon form factors can be determined from the Dalitz plot
distribution of the $K_{e2\gamma}$ decay width, 
since the term $f_\mathrm{IBSD}$ (see Eq.~\eqref{fIBSD}) is small.  
In  
$K_{\mu2\gamma}$ both the linear and quadratic terms in $F_A$ and $F_V$
contribute at comparable levels, making
it
possible to measure the signs as well as the magnitudes of the form
factors.  
Unfortunately, in the region where the linear terms grow,
the dominant contribution to $K_{\mu2\gamma}$ (the $\mathrm{IB}$ 
term which depends 
only on $f_K$) also increases,  
significantly reducing  
 the sensitivity of $K_{\mu2\gamma}$ experiments
to these form factors.  In practice, the situation is even worse,
since in these experiments only the absolute
value of the sum of the kaon form factors has been measured with good
accuracy, while their difference still
has only lower and upper bounds~\cite{Adler:2000vk,Hagiwara:fs}:
\begin{gather}
  |F_V+F_A|=0.165\pm0.013\;,
  \label{bound-on-sum-from-Kl2gamma}\\
  -0.24<F_A-F_V<0.04 \;.
  \label{bound-on-dif-from-Kl2gamma}
\end{gather}

\begin{figure*}
  \begin{center}
    \includegraphics[width=\textwidth]{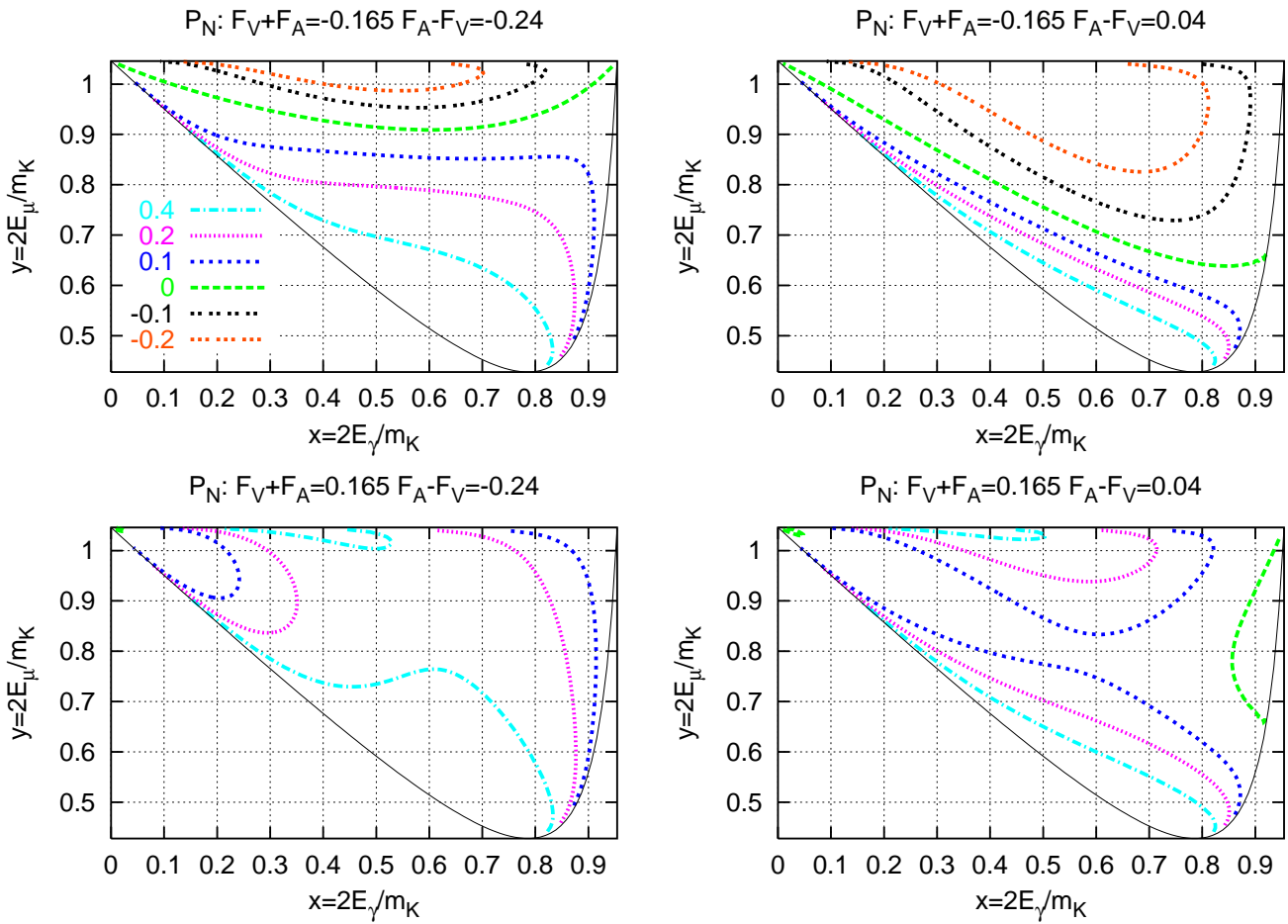}
  \end{center}
  \caption{$P_N$ distribution for different $F_V+F_A$ and $F_A-F_V$.
 In the cases of
opposite sign 
 of the sum of the form factors, 
$P_N$ also  
has opposite sign 
 at large $y$ value.
  \label{fig:Pt_Dalitz}}
\end{figure*}

The Dalitz plot distributions of $P_N$ for several values of the
form factors satisfying Eqs.~\eqref{bound-on-sum-from-Kl2gamma} and
\eqref{bound-on-dif-from-Kl2gamma} are presented in
Fig.~\ref{fig:Pt_Dalitz}. 
One can see that the  normal polarization is very sensitive to
the signs of $F_V$ and $F_A$, especially at large $y$. There in the case of
opposite signs of the sum of the form factors the $P_N$ values also  
have opposite signs.  

Recently, both vector and axial-vector form factors have been measured
in $K^+\to\mu^+\nu_{\mu}e^+e^-$ and $K^+\to e^+\nu_{e}e^+e^-$
decays~\cite{Poblaguev:2002ug}.  These decays are generalizations of
$K_{l2\gamma}$ for the case of a virtual photon in the final state, so
$F_V$ and $F_A$ are believed to be the same in all these processes.
The combined fit for both four-body decay experiments results in
\begin{equation}\label{measured-values}
  F_V=-0.112\pm0.018 \;,\qquad F_A=-0.035\pm0.020 \;.
\end{equation}
These values are in a good agreement with ${\cal O}\l p^4\r$
predictions~\cite{Hagiwara:fs,CFT} of the chiral perturbation theory (ChPT)
\begin{equation}
  F_V=-0.096 \;,\qquad F_A=-0.041 \;.
  \label{CFT-predictions}
\end{equation} 
The distributions 
 of $P_N$ over the Dalitz plot evaluated for the
measured~(\ref{measured-values}) and predicted~(\ref{CFT-predictions})
values of the form factors exhibit  
behavior similar to the upper right 
plot in Fig.~\ref{fig:Pt_Dalitz}.

Actually, form factors $F_V$ and $F_A$ are not constants, but  
functions of the momentum of the lepton pair,
$Q^2\equiv(p_K-q)^2=m_K^2(1-x)$.  In ChPT $Q^2$-dependence emerges due
to higher order corrections, which have not been calculated yet.  
These corrections generally 
decrease the chances to determine $F_V$ 
and $F_A$ in $K_{l2\gamma}$ experiments with reasonable precision. The 
$Q^2$-dependence can be estimated~\cite{CFT} supposing that the dominant
contribution comes from the exchange of the first strange hadronic
vector and axial-vector resonances with masses $m_V$ (the $K^{\ast}$ mass) 
and $m_A$ (the $K_1$ mass)
\begin{equation}\label{q2dep}
  F_{V,A}(x) = \frac{F_{V,A}}{1-\frac{m_K^2}{m_{V,A}^2}(1-x)} \;.
\end{equation}
The results from the recent $K\to l\nu e^+e^-$ 
experiment~\cite{Poblaguev:2002ug} favor this
$x$ dependence    
over constant form
factors. Unfortunately, the statistics accumulated in this experiment 
did not allow  
definite 
confirmation  or rejection of 
  $x$ dependence. 
Until now $Q^2$-dependence remains unknown theoretically 
and new experiments with higher statistics are needed to 
fix it experimentally.


For the following analysis we need to estimate the level of
statistical precision 
in the measurement of $P_N$ which can be achieved in the currently 
running and
forthcoming experiments. Generally, with the analyzing power of the detector
$\alpha$ and the kinematical attenuation factor $f$, the expected
sensitivity to $P_N$ in some region ${\cal R}$ of the Dalitz plot
can be estimated as 
\begin{equation}\label{eq:pt}
  \delta P_N({\cal R}) \simeq \frac{1}
   {\alpha f\sqrt{N_{K_{\mu2\gamma}}({\cal R})}}\;,
\end{equation}
where $N_{K_{\mu2\gamma}}({\cal R})$ is the number of
$K_{\mu2\gamma}$ events in the region ${\cal R}$. 
This should be compared to the expected value of the normal muon 
polarization 
\begin{equation}\label{eq:pn}
  P_N({\cal R})=\frac{\int_{\cal R}\rho_N(x,y)\,dx\,dy}{\int_{\cal
  R}\rho_0(x,y)\,dx\,dy}\;.      
\end{equation}

\begin{figure}
  \begin{center}
    \includegraphics[width=0.85\columnwidth]{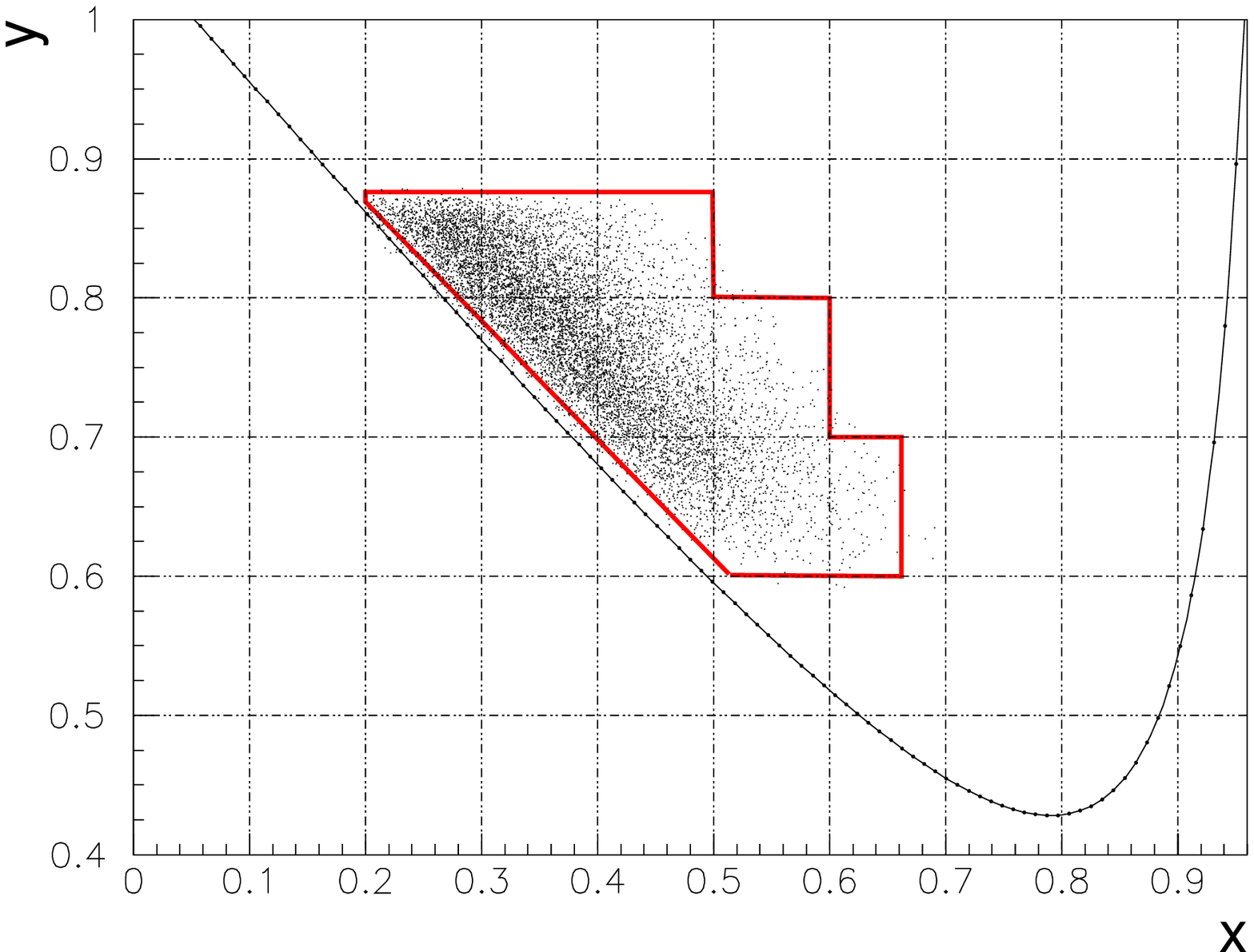}
  \end{center}
  \caption{The Dalitz plot with $K_{\mu2\gamma}$ events accumulated in
the E246 experiment~\cite{Kudenko:2000sc}.  The region of the Dalitz
plot confined by the solid lines was used for the calculation of the $P_N$
values.}
\label{fig:E246}
\end{figure}

The ongoing E246 experiment at KEK \cite{Abe:1999nc}
dedicated to a measurement of $P_T$ in the decay $K^+\to
\pi^0\mu^+\nu$ has only a limited sensitivity of about $10^{-2}$ to
both $P_T$ and $P_N$ in $K_{\mu2\gamma}$ \cite{Kudenko:2000sc}.  About
$2\times 10^5$ $K_{\mu2\gamma}$ events (see Fig.~\ref{fig:E246}) are
expected to be accumulated in the region of the Dalitz plot where the
IB term dominates.
Integrating over the region confined by thick
solid line in Fig.~\ref{fig:E246}, we obtain
\begin{gather}
P_N(F_V\!+\!F_A\!=\!0.165\;,F_A\!-\!F_V\!=\!-0.24)=0.360\;,
  \nonumber\\
P_N(F_V\!+\!F_A\!=\!0.165\;,F_A\!-\!F_V\!=\!0.04)=0.210\;,
  \nonumber\\
P_N(F_V\!+\!F_A\!=\!-0.165\;,F_A\!-\!F_V\!=\!-0.24)=0.318\;,
  \nonumber\\
P_N(F_V\!+\!F_A\!=\!-0.165\;,F_A\!-\!F_V\!=\!0.04)=0.166\;.
  \nonumber
\end{gather}
With values $\alpha\simeq 0.3$ and $f\simeq 0.65$ adopted in the
experiment, the statistical sensitivity obtained using
Eq.~(\ref{eq:pt}) is estimated to be $\sim 1.2\times10^{-2}$.  Hence,
the analysis of the E246 data will determine
the signs of the
kaon form factors
for sure if the difference between the $F_V$ and
$F_A$ values is fixed from other considerations.

The sensitivity of the E246 experiment to the form factor values can
be estimated from the difference between $P_N$ values obtained for
experimental form factors~(\ref{measured-values}) and for the ChPT
predictions~(\ref{CFT-predictions}):
\begin{gather}
P_N(F_V\!=\!-0.112\;,F_A\!=\!-0.035)=0.145\;,
\nonumber
\\
P_N(F_V\!=\!-0.096\;,F_A\!=\!-0.041)=0.161\;.
\nonumber
\end{gather}
This yields $\Delta P_N = 1.6\times 10^{-2}$ which is comparable to
the expected statistical error of E246.  One can conclude that these
values can be distinguished at the level of about 1$\sigma$, but only if 
the systematic error will be as small as 1\%. 

We note 
 in passing that the ambiguity associated with unknown
$Q^2$-dependence reduces the sensitivity of the E246 experiment to the
form factors, if $Q^2$-corrections are not fixed.  To deal with this
one can, for example, fit the experimental data with constant
$F_{V,A}$ form factors and with form factors given by Eq.~(\ref{q2dep}).
If the latter turns  
out 
 to be more favorable (as in the case of   the 
$K\to l\nu
e^+e^-$ experiment~\cite{Poblaguev:2002ug}), then  
$P_N$ should be
calculated assuming the same dependence.  In this case
\begin{gather}
P_N(F_V\!=\!-0.112\;,F_A\!=\!-0.035)=0.128\;,
\nonumber
\\
P_N(F_V\!=\!-0.096\;,F_A\!=\!-0.041)=0.147\;.
\nonumber
\end{gather}


A new experiment at Japaneese Hadron Facility (JHF)~\cite{jhf}
in which a statistical sensitivity of
$\lesssim10^{-4}$ for both $P_N$ and $P_T$ in $K_{\mu2\gamma}$ was
proposed in~\cite{Kudenko:yk}.  The main features of this experiment
include a high resolution measurement of neutral particles from
$K_{\mu3}$ and $K_{\mu2\gamma}$ decays, an active muon polarimeter
which provides information about stopped muons (stopping point,
momentum), positron direction, and also detects photons, and a highly
efficient photon veto system covering nearly $4\pi$ solid angle.  This
approach allows  
$K_{\mu2\gamma}$ events for all
$\theta$ angles between photon and muon momenta 
to be accumulated 
due to efficient
photon veto  
detection. This system eliminates $K_{\pi2}$ decays which
are estimated to be the main background source at large
$\theta$. Thus, the region of large $x$ and $y$ becomes available for
studying $P_N$.  In this experiment, the number of $K_{\mu2\gamma}$
events in the signal region with $E_{\gamma} > 20$ MeV and $E_{\mu}> 200$ MeV  
is estimated to
be about
$3\times 10^{10}$ for a 
 one year running period and beam
intensity of about $10^7 K^+/{\rm sec}$.  With  
$\alpha\simeq 0.3 $ and $f\simeq 0.8$ this would provide  
a  very low
statistical error $\delta P_T\sim 2.5\times 10^{-5}$
over  
 the entire Dalitz plot.  This experiment is expected to control the
systematic error at   
a level comparable to the statistical
uncertainty. However, in the measurement of $P_N$, where expected
non-zero values of normal asymmetry could be a few tens percent,
the accuracy will be  
diminished due to the error  
associated with the
polarimeter analyzing power and uncertainty in the level of background
present in $K_{\mu2\gamma}$ events from other kaon decays and
accidentals.  It is rather difficult to measure $P_N$
with an
error less than a few percent, and  
therefore 
this restricts the
accuracy in extraction of the form factor values.  In order to
 significantly 
improve the accuracy for extraction of $F_V$ and $F_A$
values we looked at the relative changes of $P_N$ values in several
Dalitz plot regions.

Let us evaluate the accuracy which can be obtained for $F_V$ and $F_A$
in this experiment. Changing $F_V+F_A$ and $F_A-F_V$ by 1\% and 3\%
around the ChPT predictions we examine the corresponding effect on the
normal polarization for different regions of the Dalitz plot. The
results are presented in Table~\ref{table:1}.
\begin{table*}
\caption{The $P_N$ values in several regions of Dalitz plot calculated
according to Eq.~(\ref{eq:pn}) and their sensitivity 
to small changes of the form factor values from the ChPT predictions 
 $F_V= -0.096$, $F_A= -0.041$ :\\ 
\mbox{ }$\Delta P_N^{+a}=
 P_N[(1+a\times10^{-2})(F_V\!+\!F_A),\,F_A\!-\!F_V]
-P_N[F_V\!+\!F_A,\,F_A\!-\!F_V]$;\\
\mbox{ }$\Delta P_N^{-a}=
 P_N[F_V\!+\!F_A,\,(1+a\times10^{-2})(F_A\!-\!F_V)]
-P_N[F_V\!+\!F_A,\,F_A\!-\!F_V]$,\\
where $a$ takes values 1.0 and 3.0.
The photon energy cut $E_\gamma>20$~MeV and muon energy cut $E_\mu>200$~MeV
are adopted; the values of 
$\delta P_N$ are the statistical errors estimated   according 
to Eq.~(\ref{eq:pt}) with $\alpha\simeq 0.3 $ and $f\simeq 0.8$ 
in corresponding regions of the Dalitz plot.}
~\\
 \begin{tabular}{|c|c|c|c|c|c|c|}
  
\hline
Region                               & $P_N$      & $\delta P_N \times 10^5$          
& $\Delta P_N^{+1.0}\times 10^5 $  &     $\Delta P_N^{-1.0} \times 10^5$ 
& $\Delta P_N^{+3.0}\times 10^5 $   &     $\Delta P_N^{-3.0}\times 10^5 $ \\
\hline
$y>0.9$, $0^\circ<\theta<60^\circ$   & 0.090  & $4.2 $ & -0.88  
& -3.1 & -2.6 & -9.2\\
$y>0.9$, $60^\circ<\theta<110^\circ$ & 0.002  & $4.4 $ & -8.8   & -5.8 
& -26.5 & -17.5\\
$y<0.9$, $0^\circ<\theta<60^\circ$   & 0.198   & $6.0 $ & -2.3   
&-10.6 & -6.9 & -31.9\\
$y<0.9$, $60^\circ<\theta<110^\circ$ & 0.019  & $7.4 $ & -22.2   
& -23.4 & -66.7 & -70.2\\
\hline
$y>0.9$, $110^\circ<\theta<160^\circ$ & -0.099 & $8.2 $ 
& -77.9   & -15.7 & -233.9 & -47.2\\
$y<0.9$, $110^\circ<\theta<160^\circ$ & -0.166  & $16.6 $ 
& -121.2   & -68.3 & -364.0 & -205.2\\
\hline
\end{tabular}
\label{table:1}
\end{table*}
As seen from this table, the behavior of  $P_N$ differs dramatically when 
either the sum or difference of the form factors is changed by 
1\% from its ChPT value. 
While $P_N$ is almost unchanged for Dalitz region $y>0.9$,
$0^\circ<\theta<60^\circ$, its values are very sensitive to small
changes in form factors in the regions with large $\theta$ (lines 5
and 6 in Table~\ref{table:1}). Statistical  
errors   
of $P_N$  
are small enough  
to
clearly distinguish this effect.  It looks possible that even
1\%-deviations of the form factors values from the ChPT
predictions~(\ref{CFT-predictions}) can be measured in the
experiment~\cite{Kudenko:yk}.
 
It is worth noting that the $P_N$ value depends on the ratios
$F_{V,A}/f_K$ rather than on the form factors themselves. The current
uncertainty in the $f_K$ value is also about 1\%~\cite{Hagiwara:fs},
thus the method presented above indeed permits to achieve
the same
precision in the measurement of $F_V$ and $F_A$.
 
 Finally, if the $Q^2$-dependence of the form factors is not fixed
 theoretically, it reduces the accuracy of the determination of
the form factors. Similarly to the experiment~\cite{Poblaguev:2002ug}
one can fit the data by constant form factors, by form factors
depending on $x$ as in Eq.~(\ref{q2dep}), or by form factors with
polynomial 
dependence on $x$ with unknown coefficients to be
determined from  
the best fit. Alternatively,
$Q^2$-dependence may be taken into account by extracting $F_V$ and
$F_A$ for a set of the $P_N$ values obtained by integrating over
several narrow bins in $x$. In any case 
 $Q^2$-dependence will be
fixed to some extent by one of these procedures.  This certainly
decreases the sensitivity of the measurements of the $P_N$ to the kaon
form factors.  Nevertheless, taking into account the large statistics
expected in the proposed experiment one may hope to achieve the
accuracy of a few percent in determination of $F_V$ and $F_A$.


In the analysis presented above, we have 
demonstrated that the
normal muon polarization $P_N$ in  
 $K_{\mu2\gamma}$ decay is a very
effective observable to pin down the kaon form factors $F_V$ and
$F_A$.  It was found that the distribution of the $P_N$ over the
Dalitz plot is very sensitive to the values of the kaon form factors.
The best sensitivity is exhibited in the region of large angles
$\theta$ between the   
outgoing muon and photon.  The measurement of
$P_N$ with the accuracy of about $10^{-2}$ pins down
the signs of the form factors
 while $10^{-4}$ is required in order to
 pin down  
their values also. 

The longitudinal $P_L$ and transverse $P_T$ muon polarizations are
also very sensitive to the values of the kaon form factors. $P_L$
arises at the tree level and, though its measurement with high precision
in the experiments discussed above is very difficult, could be used as
a cross check observable. $P_T$ arises only at the one-loop level
(so-called FSI contribution), thus only the signs of the form factors
could be determined for sure. In a forthcoming paper we will 
 show that   
in the proposed experiment at 
JHF, 
reasonable accuracy could be achieved with statistics 
 expected to be reached of $\gtrsim 10^{9}$ $K_{\mu2\gamma}$  
events at large $\theta$.

It should be noted that the signs of the pion form factors $F^\pi_V$
and $F^\pi_A$ also have not been measured yet~\cite{Hagiwara:fs}.  The
experimental situation there is similar to kaons, though $F^\pi_V$ has
a definite CVC prediction and the pion form factors are expected to be
almost constant over the whole Dalitz plot and $P_N$ value ranges from
0.1 to 1.  From the analysis given above, we can suggest that the
measurement of the lepton polarization in $\pi_{l2\gamma}$ might allow
the pion form factors to be pinned down.  This issue will be
considered elsewhere.

We would like to thank J.~Imazato and J.~Macdonald for valuable 
comments. The work is supported in part by RFBR grant 02-02-17398 and by the
program SCOPES of the Swiss National Science Foundation, project
No.~7SUPJ062239.  The work of F.B.\ is supported in part by CRDF grant
RP1-2364-MO-02.  The work of D.G.\ is supported in part by the INTAS
YSF 2001/2-142.


\end{document}